\documentclass[preprint,showpacs,preprintnumbers,amsmath,amssymb]{revtex4}

\usepackage{graphicx}
\usepackage{dcolumn}
\usepackage{bm}

\begin{document}

\preprint{APS/123-QED}

\title{Spectroscopy and intruder configurations of $^{33}$Mg and $^{31}$Ne \\
studied with antisymmetrized molecular dynamics}

\author{M. Kimura}
\affiliation{Creative Research Institution (CRIS), Hokkaido University, Sapporo
001-0021, Japan}%

\date{\today}

\begin{abstract}
 Excitation spectra and neutron single particle configurations of $^{33}$Mg and $^{31}$Ne
 are investigated by using antisymmetrized molecular dynamics combined with generator
 coordinate method. It is shown that both nuclei have strongly deformed  $3/2^-$ ground state
 with a  $3p2h$ configuration. The excitation spectra are qualitatively understood in terms
 of the Nilsson model and the calculation has shown the coexistence of different intruder
 configurations within small excitation energy. The calculated one neutron separation energy
 of  $^{31}$Ne is rather small ($S_n=250$ keV) and implies a $p$-wave one neutron  halo with 
 a strongly deformed core. 
\end{abstract}

\pacs{Valid PACS appear here}
\maketitle

\section{Introduction}
The breaking of neutron magic number $N=20$ that was firstly pointed out from the
observation of the anomalous binding energy and spin of $^{31}$Na \cite{Thi75,Hub78} has
been one of 
the great interests in nuclear physics. Based on the theoretical calculations, neighboring 
nuclei around $^{31}$Na was named ``island of inversion'' \cite{Pov86,War90} because they
are dominated by the intruder configuration in which neutrons are promoted into $pf$-shell 
across $N=20$ shell gap due to strong deformation \cite{Fuk92,Mot95}. Since then, numerous
experimental and theoretical studies have been devoted to the island. One of the recent
experimental finding is that there are coexistence of  ``spherical and deformed shape'' or
``normal and intruder configurations'' in many cases \cite{Tak09,Sch09,Wan10,Wim10}.  It is
expected that the  exploration of the coexistence will bring us further understanding of the
shell structure in neutron rich $N\sim 20$ region.

The spectroscopy of odd-neutron nuclei is of importance and interest, because their spin
and parity are related to the last neutron particle's or hole's orbit. For example, we have
predicted the coexistence of many-particle and many-hole configurations at small
excitation energy in $^{31}$Mg ($N=19$) \cite{Kim07}, and some of these excited states are  
identified by the the  proton knock-out reaction from $^{32}$Al \cite{Mil09}. The
spin-parity of  the excited states are associated with the  neutron orbits
$[N,n_z,l_z,j_z]=[2,0,2,3/2]$, [3,3,0,1/2], [2,0,0,1/2] and [3,2,1,3/2] in terms of the Nilsson
model. 

In this paper, we extend our study to $N=21$ system ($^{33}$Mg and $^{31}$Ne).
Both of them  do not have the definite spin-parity assignment for the ground states and
information of their excitation spectra is deficient. In the case of $^{33}$Mg, there are two
major spin-parity assignments of the ground state, $3/2^+$ and $3/2^-$. From the observation
of the $\beta$ decays of $^{33}$Na \cite{Num01} and $^{33}$Mg \cite{Tri08}, $3/2^+$
assignment is suggested. On the other hand, the measurement of the magnetic moment
\cite{Yor07} and one neutron knock-out reaction from $^{34}$Mg \cite{Kan10} suggest
$3/2^-$.  $^{31}$Ne is more interesting. Its large reaction cross section \cite{Tak10} and one 
neutron Coulomb breakup cross section \cite{Nak10} are reported. From the analysis of the
one neutron breakup, the assignments of $1/2^+$ or more likely $3/2^-$ are suggested. In  
both assignments, observed large Coulomb breakup cross section  strongly suggests one
neutron-halo structure of $^{31}$Ne that is firstly observed in the island. 

We have applied antisymmetrized molecular dynamics (AMD) with Gogny D1S force
\cite{Gog80} to 
investigate the spectra of $^{33}$Mg and $^{31}$Ne. It will be shown that  both nuclei
have the strongly deformed $3/2^-$ ground states with a neutron $3p2h$ configuration.
Similar to $N=19$ system, different particle-hole configurations coexist within small
excitation energy and their spin-parity are associated with the Nilsson-like orbits occupied
by the last neutron. It is also shown that the one neutron separation energy of $^{31}$Ne is
rather small and  $p$-shell neutron-halo structure is possibly formed. 

This paper is organized as follows. In the next section, the framework of AMD and the
calculational procedure are explained. In the section \ref{sec::results}, the results are
presented. The change of neutron particle-hole configurations as function of deformation is
discussed. They are related to the spin-parity of the excitation spectra. The calculated
spectra and transition probabilities are compared with the observations, and a theoretical
assignment of the excitation spectra is suggested. The final section summarizes this work.

\section{Theoretical Framework}\label{sec::framework}
The applied theoretical method is the same as our previous work \cite{Kim07}. The
deformed-basis AMD \cite{Kim01,Kim04-1} is combined with the generator coordinate method
(GCM) to calculate the excitation spectra of $N=21$ system. The  particle-hole configuration 
is investigated by the analysis of neutron single particle orbits. 

\subsection{Wave function, Hamiltonian and variation}
The intrinsic wave function of the system with mass A is given by a Slater determinant
of single particle wave packets,
\begin{eqnarray}
 \Phi_{ int} &=& {\mathcal A}
  \{\varphi_1,\varphi_2,...,\varphi_A \} ,\label{EQ_INTRINSIC_WF}\quad
  \varphi_i({\bf r}) = \phi_i({\bf r})\chi_i\xi_i, 
\end{eqnarray}
where $\varphi_i$ is the $i$th single particle wave packet consisting of the spatial
$\phi_i$, spin $\chi_i$ and isospin $\xi_i$ parts. The local Gaussian located at
${\bm Z}_i$ is employed as $\phi_i$, 
\begin{eqnarray}
 \phi_i({\bf r}) &=& \exp\biggl\{-\sum_{\sigma=x,y,z}\nu_\sigma
  \Bigl(r_\sigma - 
  \frac{Z_{i\sigma}}{\sqrt{\nu_\sigma}}\Bigr)^2\biggr\}, 
  \nonumber\\ 
 \chi_i &=& \alpha_i\chi_\uparrow + \beta_i\chi_\downarrow,
  \quad |\alpha_i|^2 + |\beta_i|^2 = 1,\nonumber \\
 \xi_i &=& {\rm proton} \quad {\rm or} \quad {\rm neutron}. \label{EQ_SINGLE_WF}
\end{eqnarray}
Here ${\bm Z}_i$, $\alpha_i$, $\beta_i$ and $\nu_\sigma$ are the variational
parameters. The parity projected wave function, $\Phi^{\pi} = \hat{P}^\pi \Phi_{ int}$
is the variational wave function. 

The Gogny D1S force  is employed as an effective nuclear force and
the Coulomb force is approximated by a sum of seven Gaussians.   

In the present work, we have performed the variational calculation with the constraint on
the matter quadrupole deformation $\beta$. Variational parameters are optimized by the
frictional cooling method so that the energy of the system is minimized for a given
constraint on $\beta$. The optimized wave function are denoted as $\Phi_{int}(\beta)$. It is
noted that we do not put any constraint on the matter quadrupole deformation
$\gamma$. Therefore $\gamma$ has the optimal value for each $\beta$.

\subsection{Analysis of the single particle orbit}
To identify the particle-hole configuration, we analyze the neutron single particle
orbital of $\Phi_{int}(\beta)$. We transform the  single  particle wave packet $\varphi_i$
to the orthonormalized basis,    $\widetilde{\varphi}_\alpha = 
  \frac{1}{\sqrt{\lambda_\alpha}}\sum_{i=1}^{A}c_{i\alpha}\varphi_i.$
Here, $\lambda_\alpha$ and $c_{i\alpha}$ are the eigenvalues and eigenvectors of the
overlap matrix $B_{ij}=\langle\varphi_i|\varphi_j\rangle$. Using this basis, the
 Hartree-Fock single particle Hamiltonian,   
\begin{eqnarray}
 h_{\alpha\beta} &=&
  \langle\widetilde{\varphi}_\alpha|\hat{t}|\widetilde{\varphi}_b\rangle + 
  \sum_{\gamma=1}^{A}\langle
  \widetilde{\varphi}_\alpha\widetilde{\varphi}_\gamma|
  {\hat{v}_n+\hat{v}_c}|
  \widetilde{\varphi}_\beta
\widetilde{\varphi}_\gamma -
\widetilde{\varphi}_\gamma\widetilde{\varphi}_\beta\rangle,\nonumber\\ 
&+&\frac{1}{2}\sum_{\gamma,\delta=1}^{A}
 \langle\widetilde{\varphi}_\gamma\widetilde{\varphi}_\delta 
|\widetilde{\varphi}_\alpha^*\widetilde{\varphi}_\beta
\frac{\partial\hat{v}_n}{\partial \rho}|\widetilde{\varphi}_\gamma
\widetilde{\varphi}_\delta - \widetilde{\varphi}_\delta  \widetilde{\varphi}_\gamma
\rangle, 
\end{eqnarray}
is defined. The eigenvalues $\epsilon_s$ and eigenvectors  $f_{\alpha s}$ of
$h_{\alpha\beta}$ give the single particle energies and the single particle orbits, 
$\widetilde{\phi}_s = \sum_{\alpha=1}^{A}f_{\alpha s}\widetilde{\varphi}_\alpha$.
This procedure gives us information of the occupied neutron orbits. In this paper the 
particle-hole configuration is denoted as $mpnh$ where $m$ and $n$ respectively denote
number of neutron particles in $pf$-shell and holes in $sd$-shell in the spherical
limit. For example, the normal configuration of $N=21$ system is denoted as $1p0h$. 
However, it is noted that this procedure only gives a rough estimation of the particle-hole
configuration and should be regarded as a guide to understand the dominant particle-hole
configuration of each states. The actual variational wave function is $\Phi^\pi(\beta)$ that
is a sum of two Slater determinant. And we then perform the angular momentum projection and
GCM calculation that superpose thousands of Slater determinants. Therefore our final wave
function explained in the next subsection is beyond a simple single particle picture.

\subsection{Angular momentum projection and GCM}
After the variational calculation, we project out an eigenstate of the total angular
momentum $J$,  
\begin{eqnarray}
 \Phi^{J\pm}_{MK}(\beta) = \hat{P}^{J}_{MK}\Phi^{\pm}(\beta)
  \label{EQ_ANGULAR_WF}
\end{eqnarray} 
Here $\hat{P}^{J}_{MK}$ is the total angular momentum projector. The integrals over
three Euler angles are evaluated by the numerical integration. The calculation is completed
by performing GCM. The wave functions $\Phi^{J\pm}_{MK}(\beta)$ which have the same
parity and  angular momentum but have different  $K$, deformation $\beta$ and single
particle configurations are superposed. Then the wave function of the system is expressed as 
\begin{eqnarray}
 \Phi_n^{J\pm} = c_n\Phi^{J\pm}_{MK}(\beta)
  + c_n^\prime\Phi^{J\pm}_{MK^\prime}(\beta^\prime) + \cdots,
  \label{eq::gcm_wf}
\end{eqnarray}
where the quantum numbers except for the total angular momentum and the parity are
represented by $n$. The coefficients $c_n$, $c'_n$,... are determined by the
Hill-Wheeler equation. The physical quantities discussed in the next section are basically
calculated from the GCM wave function, Eq. (\ref{eq::gcm_wf}).

\section{Results and discussions}\label{sec::results}

\subsection{Energy curve and single particle configurations of $^{\bm {33}}$Mg}
\begin{figure*}[bt!]
 \includegraphics[width=\hsize]{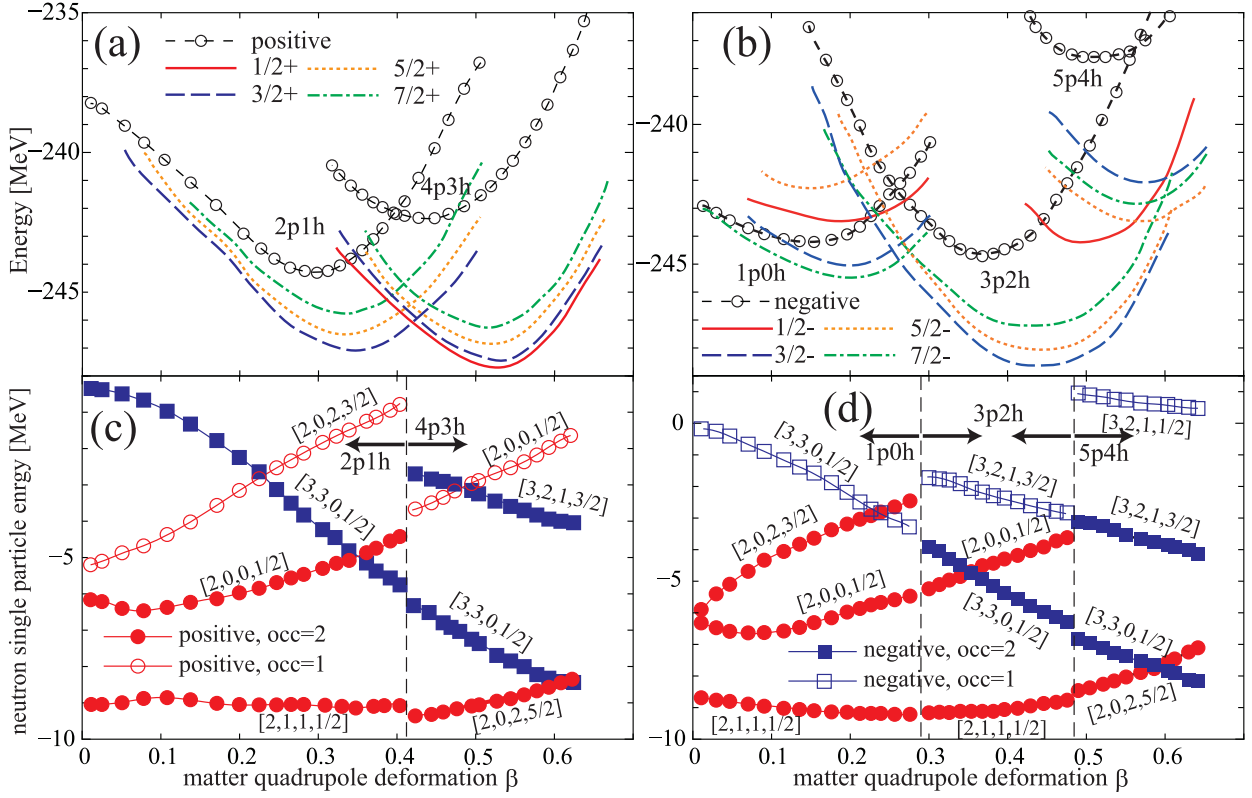}
\caption{(color online) (a) and (b): Energy curves of $^{33}$Mg before and after angular
 momentum   projection for (a) positive- and (b) negative-parity states. Circles with dashed
 lines show  the energy curve before the angular momentum projection. Colored lines show the
 energy  curves after the angular momentum projection to $J=1/2$, 3/2, 5/2 and 7/2.
 (c) and (d): Corresponding single particle energies of the last 7 neutrons. Filled (open)
 symbols show the orbits occupied by two (one) neutrons. Circles (boxes) show the orbits in
 which positive-parity (negative-parity) component is  larger than 50 \%.}
\label{fig::Mg33_curve}
\end{figure*}
\begin{figure}[hbt]
 \includegraphics[width=\hsize]{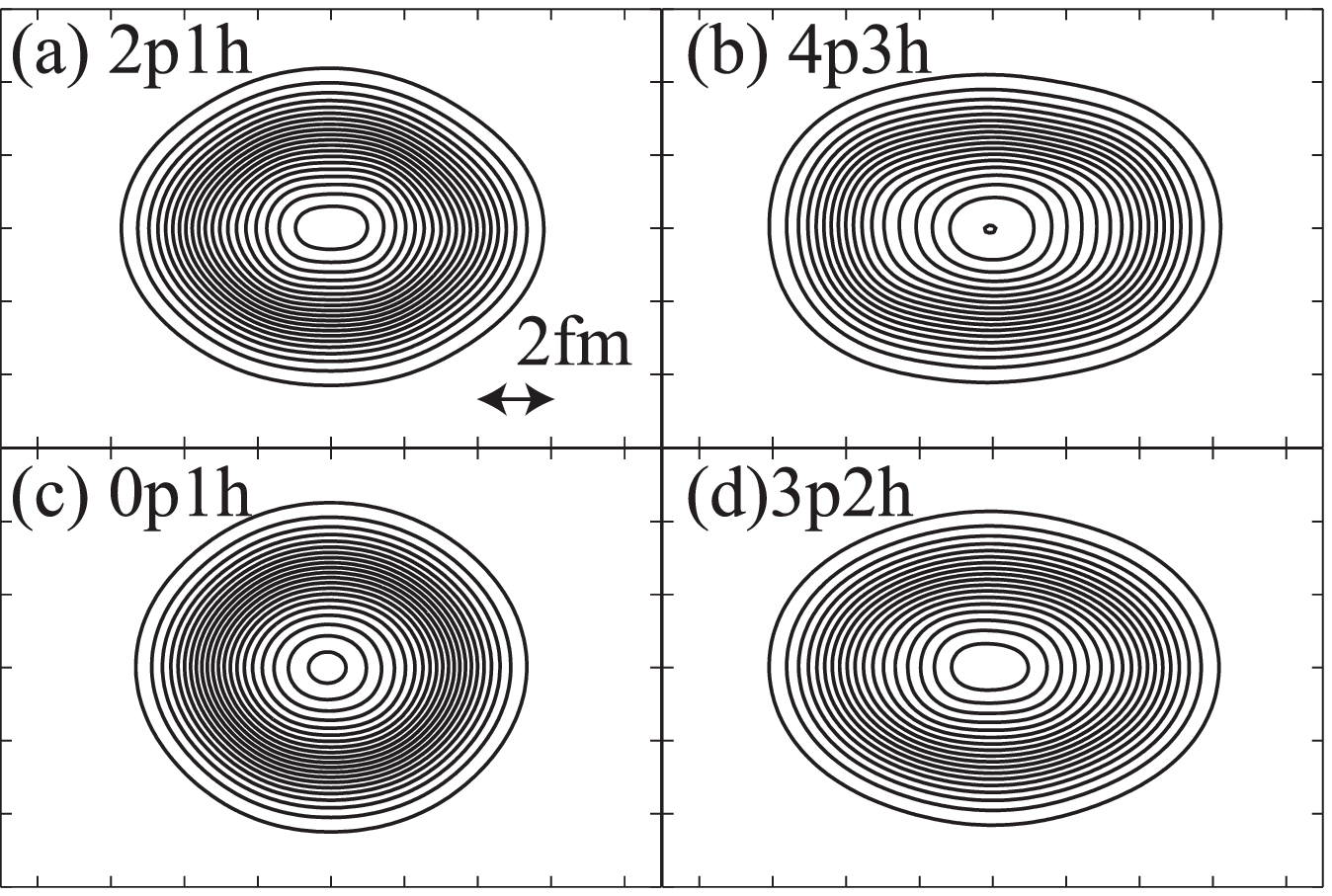}
\caption{Intrinsic density distributions at minima of energy curves. (a) and (b) show the
 minima of positive-parity states. (c) and (d) show $0p1h$ and $3p2h$ minima of
 negative-parity states.} 
\label{fig::Mg33_density}
\end{figure}

Energy curves obtained by the variational calculation after the parity projection are
presented in Fig. \ref{fig::Mg33_curve} (a) and (b) with circles. For a given value of
$\beta$, there is sometimes a local minimum solution together with the global minimum. In
such case, both minima are plotted for the same value of $\beta$. For any value of $\beta$,
$^{33}$Mg prefered prolate deformation $(\gamma > 30^\circ)$ over oblate deformation 
$(\gamma < 30^\circ)$.   The positive-parity states have two minima and they have neutron
$2p1h$ and $4p3h$ configurations in the ascending order of deformation.  The negative-parity
states have three minima with $1p0h$, $3p2h$ and   $5p4h$ configurations. The particle-hole
configurations of those minima are confirmed by the analysis of the neutron single particle
orbits. Figure \ref{fig::Mg33_curve} (c) and (d) shows the single particle energies of the
last 7 neutrons as function of  deformation. In the case of the negative parity, the last
neutron occupies $f_{7/2}$ orbit and other neutrons fill $sd$-shell at small
deformation. Therefore, the minimum around $\beta=0.2$ in  Fig. \ref{fig::Mg33_curve} (b)
has a neutron $1p0h$ configuration. As deformation becomes  larger, neutron single particle
energies show Nilsson model like behavior. Two orbits originate in $f_{7/2}$ come down and
$d_{3/2}$ splits into two orbits and go up.  They are respectively denoted as
$[N,n_z,l_z,j_z]=[3,3,0,1/2]$, [3,2,1,3/2], [2,0,2,3/2] and  [2,0,0,1/2]  in terms of
Nilsson model. Around $\beta=0.25$ the configuration changes. Two neutrons occupy
[3,3,0,1/2] and one occupies [3,2,1,3/2] orbit. The [2,0,2,3/2] orbit is now empty. Thus the
second minimum of the energy curve has a $3p2h$ configuration. Further increase of
deformation pulls down the [3,2,1,1/2] orbit originates  in $p_{3/2}$. The last neutron
occupies it and two orbits originate in $d_{3/2}$ become empty. Therefore third minimum
that appears as the local minimum above the $3p2h$ configuration has a $5p4h$
configuration. Its energy is rather higher than other two configurations. The
positive-parity curve is also understood in the same way. It has  two minima that have
respectively $2p1h$ and $4p3h$ configurations. The corresponding intrinsic matter density
distributions are shown in Fig. \ref{fig::Mg33_density}. It shows an almost spherical shape
of the normal configuration $(1p0h)$ and deformation becomes larger as number of neutrons in
$pf$-shell increases. 

The energy curves and corresponding neutron single particle orbits show the qualitative
agreement with the Nilsson model picture discussed in Ref. \cite{Ham07,Ham10}. The neutron
single particle orbits split and change their order depending on deformation of the system. As
deformation becomes larger, number of particle in the orbits originate in $pf$-shell
increases. However, it is reminded that the analysis of the neutron single particle orbits
gives only a rough estimation of the particle-hole configuration. Indeed, there are strong
parity mixing of the neutron single particle orbits. For example, the orbit denoted as
[3,2,1,1/2] in Fig. \ref{fig::Mg33_curve} (d) has the strongest parity mixing where the
the positive-parity component amounts to about 40\%. The parity mixing becomes
stronger as deformation larger. This may mean that the strongly deformed states have a
complicated mixing of different particle-hole configurations and are beyond the Nilsson
model picture. 

Adding to the Nilsson model like behavior, the present calculation has shown the
following two points. (1) $^{33}$Mg is strongly deformed and quite soft against
deformation. The intruder $3p2h$ configuration is already the lowest energy configuration
at this stage and other particle-hole configurations also appear within very small
excitation energy. (2) The last neutron walks around the [3,3,0,1/2], [2,0,2,3/2],
[3,2,1,3/2], [2,0,0,1/2] and [3,2,1,1/2] orbits as deformation becomes larger. These neutron
orbits  will characterize the spin-parity of the corresponding excited states or rotational
bands.  

Lines in Fig. \ref{fig::Mg33_curve} show the energy curves after the angular momentum 
projection. The energies of strongly deformed states are lowered by more than 5 MeV.
The lowest energy state at this stage is the $3/2^-$ state with the $3p2h$
configuration. More importantly, the property of the last neutron orbital is reflected
to the excitation spectrum of each configuration. Deformation of  $1p0h$ configuration is
the smallest among all minima and it generates low-lying $7/2^-$ and $3/2^-$ states that are
respectively attributed to the almost spherical neutron $f_{7/2}$ and $p_{3/2}$ orbits. In
the case of other minima, rotational spectra appear depending on the last neutron orbit. For
example, the $3p2h$ configuration has the last neutron in the [3,2,1,3/2] orbit and it
generates the $K^\pi=3/2^-$ rotational spectrum. In the same way, the $2p1h$, $4p3h$ and
$5p3h$  configurations respectively generate the $K^\pi=3/2^+$, $1/2^+$ and $1/2^-$ spectra
that are  associated with the asymptotic quantum number $j_z$ of the last neutron.

\subsection{Spectrum of $^{33}$Mg and transitions}
\begin{figure}[hbt] 
 \includegraphics[width=\hsize]{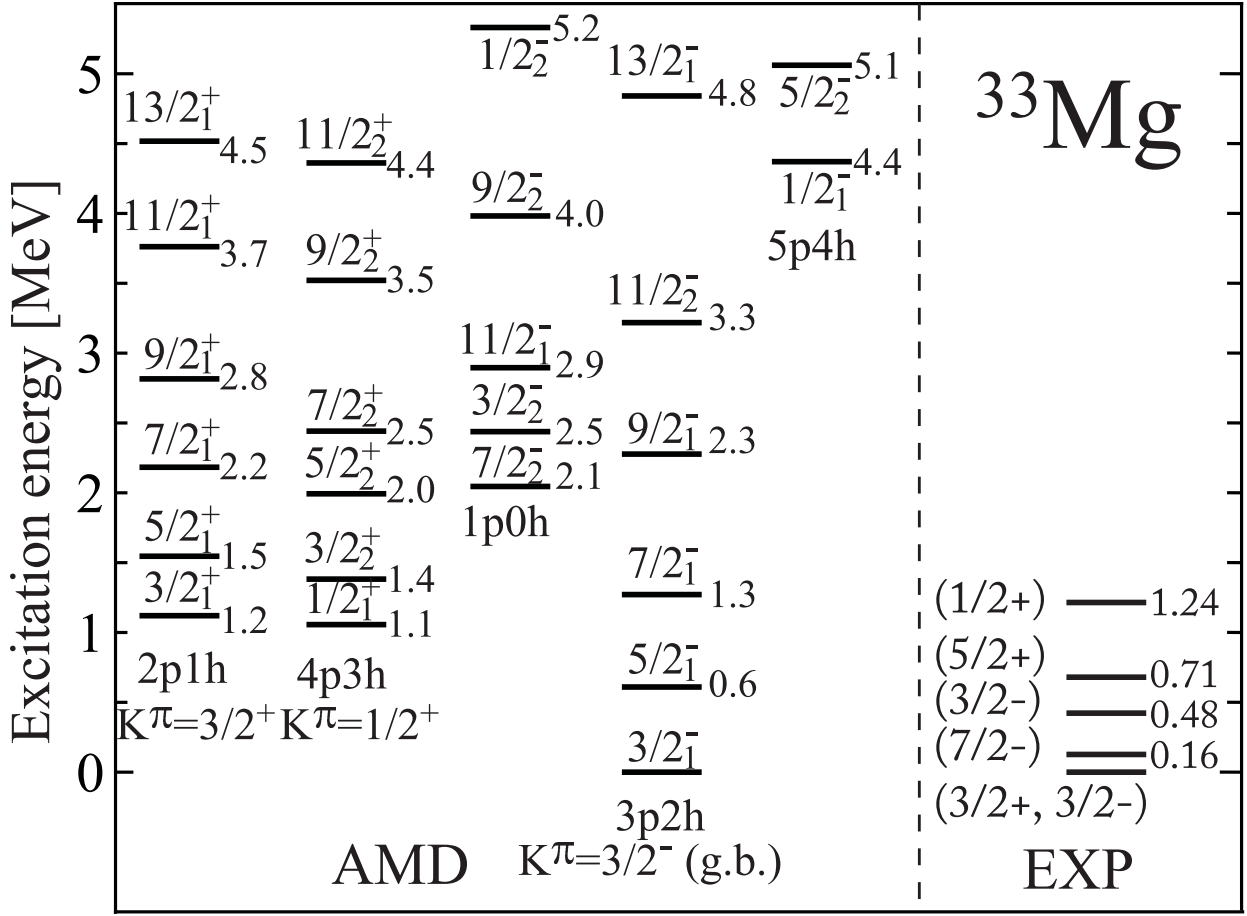}
\caption{Calculated spectrum of $^{33}$Mg compared with the tentative assignments
 suggested in Refs. \cite{Num01,Yor07}. $mpnh$ configurations shown below each bands mean
 the  dominant particle configuration. Numbers show excitation energies in MeV. } 
\label{fig::Mg33_level}
\end{figure}
\begin{figure*}[hbt] 
 \includegraphics[width=\hsize]{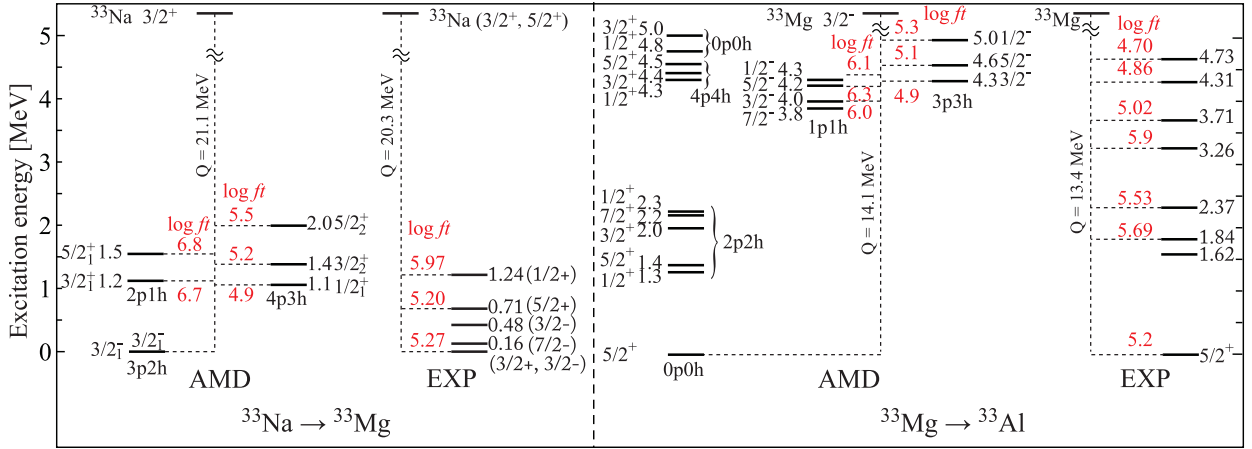}
\caption{(color online) Calculated and observed \cite{Num01,Tri08} $\beta$ decay strength of
 $^{33}$Na $\rightarrow$ $^{33}$Mg (left) and  $^{33}$Mg $\rightarrow$ $^{33}$Al (right). 
} 
\label{fig::Mg33_beta}
\end{figure*}

\begin{table}[thb]
\caption{Calculated intra- and inter-band $E2$ transition probabilities in $e^2\rm
 fm^4$}\label{tab::BE2} 
\begin{ruledtabular}
\begin{tabular}{cccc}
 band&$J^\pi_i$&$J^\pi_f$&$B(E2;J^\pi_i\rightarrow J^\pi_f)$\\
 \colrule
 g.b. $\rightarrow$ g.b.&$3/2^-_1$&$5/2^-_1$&282\\
 &$3/2^-_1$&$7/2^-_1$&154\\
 &$5/2^-_1$&$7/2^-_1$&147\\
 &$5/2^-_1$&$9/2^-_1$&194\\
 &$7/2^-_1$&$9/2^-_1$&92\\
 \colrule
 $K^\pi=1/2^+ \rightarrow 1/2^+$&$1/2^+_1$&$3/2^+_2$&275\\
 &$1/2^+_1$&$5/2^+_2$&426\\
 &$3/2^+_2$&$5/2^+_2$&66\\
 &$3/2^+_2$&$7/2^+_2$&357\\
 &$5/2^+_2$&$7/2^+_2$&24\\
 \colrule
 $K^\pi=3/2^+ \rightarrow 3/2^+$ &$3/2^+_1$&$5/2^+_1$&188\\
 &$3/2^+_1$&$7/2^+_1$&103\\
 &$5/2^+_1$&$7/2^+_1$&99\\
 \colrule
 $1p0h \rightarrow 1p0h$  &$7/2^-_2$&$3/2^-_2$&31\\
&$7/2^-_2$&$11/2^-_2$&23\\
 \colrule
 $K^\pi=1/2^+ \rightarrow 3/2^+$ &$1/2^+_1$&$3/2^+_1$&11\\
 &$1/2^+_1$&$5/2^+_2$&19\\
 &$3/2^+_2$&$5/2^+_1$&18\\
 &$3/2^+_2$&$7/2^+_1$&30\\
\end{tabular}
\end{ruledtabular}
\end{table}

\begin{table}[thb]
\caption{Calculated and observed \cite{Yor07} magnetic moments in the unit of
 $\mu_N$. Theoretical values are shown for the lowest energy states of particle-hole
 configurations.} 
 \label{tab::mag}
\begin{ruledtabular}
\begin{tabular}{cccccc}
 &$3/2^-_1 (3p2h)$&$7/2^-_1 (1p0h)$&$3/2^+_1 (2p1h)$&$1/2^+_1 (4p3h)$&exp.\\
 \colrule
 $\mu_N$&-0.70&-1.61 &0.89&-0.91&-0.7456\\
\end{tabular}
\end{ruledtabular}
\end{table}

Figure. \ref{fig::Mg33_level} shows the low-lying spectrum of $^{33}$Mg obtained by
GCM. Tentative assignments given in Refs. \cite{Num01,Yor07} are also shown. The calculated
binding energy of $^{33}$Mg is 251.1 MeV that underestimates the observed value by about 1
MeV. The one neutron separation energy is 1.9 MeV that is not far from the observed value
(2.22 MeV). The ground state is the $3/2^-_1$ state and dominated by the $3p2h$
configuration that  approximately amounts to 87\%. Here  the amount of the $3p2h$
configuration in the ground state is estimated by calculating the overlap between the ground
state GCM wave function ($\Phi^{3/2-}_{g.s.}$) and the wave function of $3/2^-$ minimum
($\Phi^{3/2^-}(\beta=0.45)$) in Fig. \ref{fig::Mg33_curve} (b),    
\begin{eqnarray}
| \langle \Phi^{3/2-}_{g.s.}|\Phi^{3/2^-}(\beta=0.45) \rangle|^2.
\end{eqnarray}
The $3p2h$ configuration dominates the ground band $K^\pi=3/2^-$ that has the first
excited state $5/2^-_1$ at 0.6 MeV. The normal configuration $1p0h$ is located at higher
excitation energy. The $7/2^-_2$ and $3/2^-_2$ states at 2.1 and 2.5 MeV are understood
as the almost spherical $f_{7/2}$ and $p_{3/2}$ single particle states.  The $2p1h$ and
$3p2h$ configurations appear around  1 MeV. They respectively dominate the $K^\pi=3/2^+$
and $1/2^+$ rotational bands. Similar to $^{31}$Mg, many-particle and many-hole states
with different deformation coexist within rather small excitation energy. 

The calculated $B(E2)$  values are summarized in Table. \ref{tab::BE2} for
several low-lying states. Except for the $1p0h$ configuration, the intra-band $E2$
transitions are large due to their deformation. Because of the mixing between different
particle-hole configurations, there are non negligible inter-band $E2$ transitions between
the $K^\pi=3/2^+$ and $1/2^+$ bands. The $\beta$ decay strength from $^{33}$Na and to
$^{33}$Al are shown in Fig. \ref{fig::Mg33_beta}. The wave functions of $^{33}$Na and
$^{33}$Al are calculated by using the same method as $^{33}$Mg. The ground state of
$^{33}$Na is $3/2^+$ and dominated by a neutron $4p2h$ configuration and mixed with 
a $2p0h$ configuration (10\%) in the present calculation. Therefore the $\beta$
decay strongly feeds the $4p3h$ configuration of $^{33}$Mg. The $\beta$ decay to $^{33}$Al
mainly feeds the $3p3h$ configuration, since the ground state of $^{33}$Mg is dominated by a
$3p2h$ configuration. It feeds the $3/2^-$, $5/2^-$ and $1/2^-$ states around 4.5 MeV
strongly.  

Then we compare our results with observations and examine the spectrum of $^{33}$Mg.  Our
result supports the  $3/2^-$ assignment of the ground state suggested from the magnetic
moment and two neutron knock-out reaction. Indeed the calculated $3/2^-_1$ state gives the
closest value to the observed magnetic moment among the lowest energy states of
particle-hole configurations (Table. \ref{tab::mag}). 
The calculated first excited state ($5/2^-_1$) at 0.6 MeV may  correspond to the observed
0.48 MeV state, because the large Coulomb excitation cross section from the ground state to
this state is reported  and large quadrupole deformation $\beta_C=0.52$ is extracted
\cite{Pri02}.  Experimentally, the first excited state was reported at 0.16 MeV from the
$\beta$ decay measurement of $^{33}$Na.  Since there was no direct $\beta$ decay feeding of
this state, a tentative assignment of $7/2^-$ was given. In our calculation, there is not
corresponding state with very small excitation energy. The observed 0.71 and 1.24 MeV states
could be associated with the calculated $1/2^+_1$ (1.1 MeV) and $3/2^+_2$ (1.4 MeV) states
of $K^\pi=1/2^+$ band from their $\log ft$ values. However, the present calculation does
not explain the $\beta$ decay of $^{33}$Na to the ground state of $^{33}$Mg nor that of
$^{33}$Mg to $^{33}$Al. Especially, our calculation has no $\beta$ decay feeding of
low-lying states of $^{33}$Al below 4 MeV, while there are 5 transitions including the
feeding of the ground state of $^{33}$Al are reported. To explain those transitions, we need
the positive-parity ground state of $^{33}$Mg.

An alternative assignment for the ground state of $^{33}$Mg could be $1/2^+_1$ with the
$4p3h$ configuration, since its magnetic moment has the same sign with and is not far from
the observed value. Though the $1/2^+_1$ state is located at 1.1 MeV in the present
calculation, a minor modification of the Gogny force parameter might make it the ground
state.  In this case,  the observed $\beta$ decay data of $^{33}$Na could be
compatible. The observed 0.71 and 1.24 MeV states could be associated with the $3/2_2^+$
and $5/2^+_2$ states at 1.4 and 2.0 MeV from their $\log ft$ values. Their excitation
energies measured from the $1/2^+_1$ state are 0.4 and 1.0 MeV. The observed 0.16 and 0.48
MeV states might be associated with the negative-parity states with $1p0h$ or $3p2h$
configurations. However, even in this scenario, we cannot explain the $\beta$ decay of
$^{33}$Mg. The $1/2^+_1$ state with $4p3h$ configuration would strongly feed the states
of $^{33}$Al  dominated by $4p4h$ configuration that are located around 4.4 MeV in the
present calculation. These transitions could be associated with the observed strong decays
to 3.71, 4.31 and 4.73 MeV states. The problem is that this assignment neither explains the
feeding of the states below 4 MeV. There are only the excited states with a $2p2h$
configuration and the ground state with the $0p0h$ configurations below 4 MeV. Since there
is always the mixing between different particle-hole configuration, the $1/2^+_1$ ground
state of $^{33}$Mg could feed these states, but it will never explain the strong feeding of
the ground state of $^{33}$Al that is dominated by $0p0h$ configuration.  

Thus the present calculation partially explains some of observations but is incompatible with
some in any interpretation. Further detailed study including the spectra of $^{33}$Na and
$^{33}$Al is required to settle down this problem, that will be done in near future.

\subsection{Spectrum of $^{31}$Ne}
\begin{figure}[hbt]
 \includegraphics[width=\hsize]{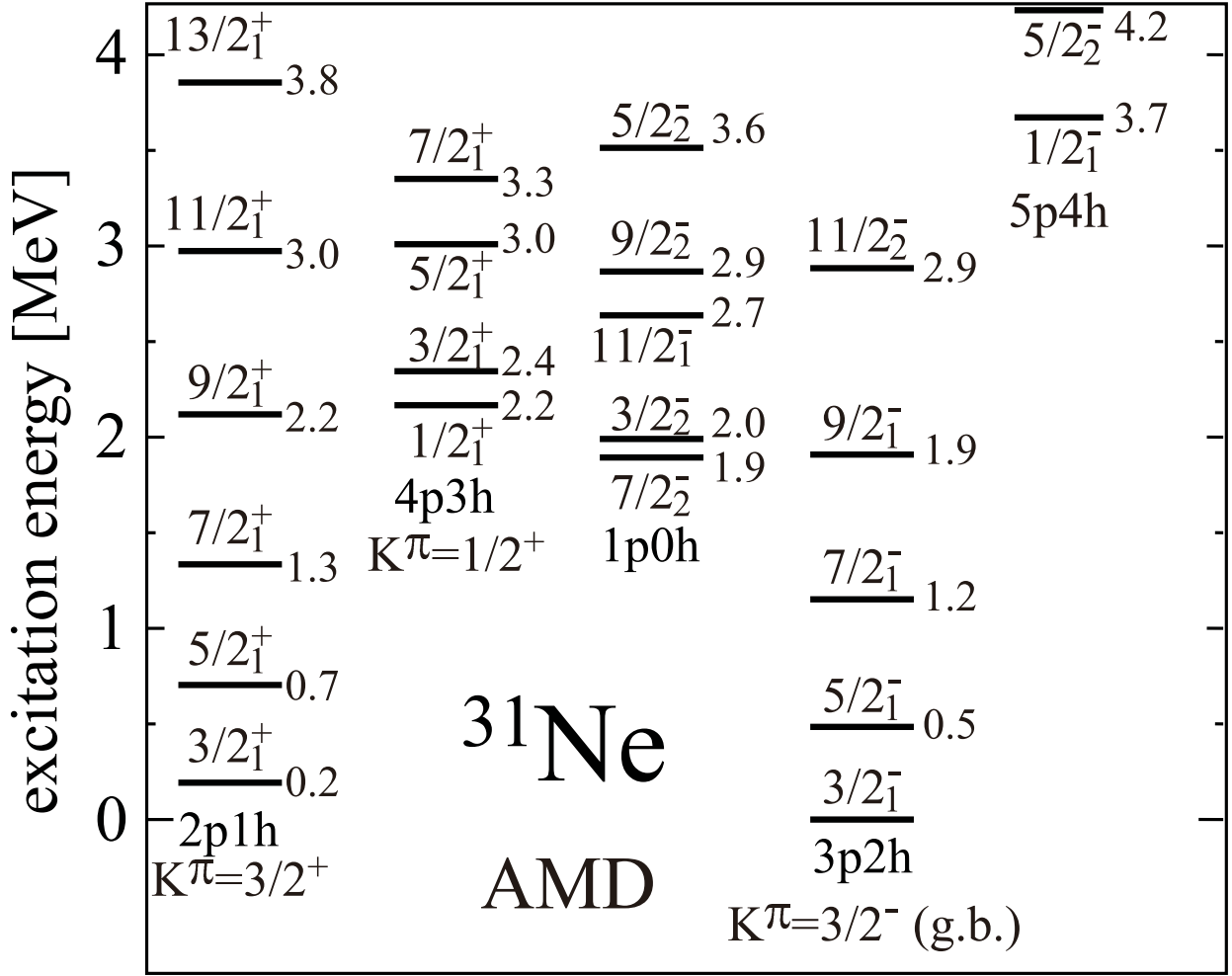}
\caption{Calculated spectrum of $^{31}$Ne. $mpnh$ configurations shown below each bands
 mean  the  dominant particle configuration. Numbers show excitation energies in MeV. }
\label{fig::Ne31_level}
\end{figure}
Since neutron number is the same, the energy curves and particle-hole configurations of 
$^{31}$Ne are qualitatively similar to $^{33}$Mg.  The difference is in the relative
energies between particle-hole configurations. The calculated energy spectrum is shown in
Fig. \ref{fig::Ne31_level}. The ground state is the $3/2^-_1$ state dominated by the $3p2h$
configuration.  Different from $^{33}$Mg, the $K^\pi=3/2^+$ band ($2p1h$) is located at
smaller excitation energy and the $K^\pi=1/2^+$ band is at higher excitation energy. We
conjecture that their excitation energies are sensitive to the neutron excess, since we have
found a minor modification of the isospin dependence of Gogny force strongly affects their
energies. 

Experimentally, small one neutron separation energy $S_n=0.29 \pm 1.64$ MeV is reported
\cite{Jur07}.
But neither the ground state spin-parity nor excitation spectrum are not well
established. The assignment of ground state $3/2^-$ or $1/2^+$ is suggested from the Coulomb  
breakup reaction \cite{Nak10}. From these observations, the $p$-wave or $s$-wave neutron
halo structure is suggested and theoretically investigated \cite{Hor10}.  Therefore, we
focus on the ground state property of $^{31}$Ne. The calculated one neutron separation
energy is $S_n=250$ keV that is fairly smaller than $^{33}$Mg. Combined with the calculated  
ground spin-parity of $3/2^-$, $^{31}$Ne is a promising  candidate of a $p$-wave halo
nucleus as suggested in Ref. \cite{Nak10}. The ground state of $^{31}$Ne can be
approximately 
regarded as the two body system consists of a $^{30}$Ne core with $2p2h$ configuration and a 
neutron in [3,2,1,3/2] orbit. Since the ground band of $^{30}$Ne is dominated by $2p2h$
configuration \cite{Kim02, Kim04-2}, there should be the collective excitation of $^{30}$Ne
core in $^{31}$Ne. The calculated matter root mean square radius of $^{31}$Ne is 3.42 fm
that is larger than $^{30}$Ne (3.31 fm), $^{32}$Ne (3.40 fm)  and $^{33}$Mg (3.34 fm). 
However, the neutron density distribution does not have a halo-like tail. This is due to the
limitation of the present calculation. Since the single particle wave function is limited to
the Gaussian form,  it cannot describe a halo tail. Improvement of the neutron wave function
by combining the AMD with the resonating group method will be made in our next study.  

\section{Summary}
In summary, we have studied the spectra of $^{33}$Mg and $^{31}$Ne. Both nuclei have the
strongly deformed $3/2^-$ ground state with a $3p2h$ configuration and the coexistence of
many different particle-hole configurations within small excitation energy. Their spectra are
qualitatively understood in terms of the Nilsson model. The last neutron occupies
$[N,n_z,l_z,j_z]=[3,3,0,1/2]$, [2,0,2,3/2], [3,2,1,3/2], [2,0,0,1/2] and $[3,2,1,1/2]$
depending on deformation of the system and the corresponding excited states or rotational
bands are formed. The calculated spectrum of $^{33}$Mg is examined based on the comparison
with the observed magnetic moment and $\beta$ decays. The $3/2^-$ assignment for the ground
state by the present calculation agrees with the observed magnetic moment, but does not
explain the $\beta$ decays. An alternative assignment of $1/2^+$ is also examined, but 
not compatible with the observed $^{33}$Mg $\rightarrow$ $^{33}$Al (g.s.) decay. This
requires further detailed study of the $A=33$ isobars. $^{31}$Ne also has the $3/2^-$ ground
state. The calculated small one neutron separation energy (250 keV) supports the $p$-wave
neutron halo structure.

\bibliography{apssamp}

\begin{references}
 \bibitem{Thi75} C. Thibault {\it et al}., Phys. Rev. {\bf C 12}, 644 (1975).
 \bibitem{Hub78} G. Huber {\it et al}., Phys. Rev. C {\bf 18}, 2342 (1978).
 \bibitem{Pov86} A. Poves and J. Retamosa, PHys. Lett. B {\bf 184}, 311 (1986).
 \bibitem{War90} E. K. Warburton, J. A. Becker and B. A. Brown,
 Phys. Rev. C {\bf 41}, 1147 (1990). 
 \bibitem{Fuk92} N. Fukunishi, T. Otsuka and T. Sebe, Phys. Lett. B {\bf 296}, 279 (1992). 
 \bibitem{Mot95} T. Motobayashi {\it et al}., Phys. Lett. B {\bf 346}, 9 (1995).
 \bibitem{Tak09} S. Takeuchi {\it et al}., Phys. Rev. C {\bf 79}, 054319 (2009).
 \bibitem{Sch09} W. Schwerdtfeger {\it et al}., Phys. Rev. Lett.  {\bf 103}, 012501 (2009).
 \bibitem{Wan10} Z. M. Wang {\it et al}., Phys. Rev. C {\bf 81}, 064301 (2010).
 \bibitem{Wim10} K. Wimmer {\it et al}., Phys. Rev. Lett.  {\bf 105}, 252501 (2010).
 \bibitem{Kim07} M. Kimura, Phys. Rev. C {\bf 75}, 041302 (2007).
 \bibitem{Mil09} D. Miller {\it et al}., Phys. Rev. C {\bf 79}, 054306 (2009).
 \bibitem{Num01} S. Nummela {\it et al}., Phys. Rev. C {\bf 64}, 054313 (2001).
 \bibitem{Tri08} V. Tripathi {\it et al}., Phys. Rev. Lett.  {\bf 101}, 142504 (2008).
 \bibitem{Yor07} D. T. Yordanov {\it et al}., Phys. Rev. Lett.  {\bf 99}, 212501 (2007).
 \bibitem{Kan10} R. Kanungo {\it et al}., Phys. Lett. B {\bf 685}, 253 (2010).
 \bibitem{Tak10} M. Takechi {\it et al}., Nucl. Phys. A {\bf 834}, 412c (2010).
 \bibitem{Nak10} T. Nakamura {\it et al}., Phys. Rev. Lett.  {\bf 103}, 262501 (2010).
 \bibitem{Gog80} J. Decharg\'e and D. Gogny, Phys. Rev. C {\bf 21}, 1568 (1980).
 \bibitem{Kim01} M. Kimura, Y. Sugawa and H. Horiuchi, Prog. Theor. Phys. (Kyoto)
  {\bf 106}, 1153 (2001).
 \bibitem{Kim04-1} M. Kimura, Phys. Rev. C {\bf 69}, 044319 (2004).
 \bibitem{Ham07} I. Hamamoto, Phys. Rev. C {\bf 76}, 054319 (2007)
 \bibitem{Ham10} I. Hamamoto, Phys. Rev. C {\bf 81}, 021304(R) (2010).
 \bibitem{Pri02} B. V. Pritychenko {\it et al}., Phys. Rev. C {\bf 65},  061304(R) (2002).
 \bibitem{Jur07} B. Jurado {\it et al}., Phys. Lett. B {\bf 649}, 43 (2007).
 \bibitem{Hor10} W. Horiuchi, Y. Suzuki, P. Capel and D. Baye, Phys. Rev. C {\bf 81}, 
 024606 (2010).
 \bibitem{Kim02} M. Kimura and H. Horiuchi, 
 Prog. Theor. Phys. (Kyoto)  {\bf 107}, 33 (2002). 
 \bibitem{Kim04-2} M. Kimura and H. Horiuchi, 
 Prog. Theor. Phys. (Kyoto)  {\bf 111}, 841 (2004). 
\end{references}

\end{document}